\documentclass[aps,twocolumn,lengthcheck,10pt,prl]{revtex4-1}

\usepackage{color}
\usepackage{pifont}
\usepackage{hyperref}
\usepackage{graphicx,epstopdf}
\usepackage{textcomp}
\usepackage{tabularx}
\usepackage{booktabs}
\usepackage{array}
\usepackage{bbm}
\usepackage{bm}
\usepackage{amssymb,amsmath}
\usepackage{fix-cm}
\usepackage[all]{xy}
\usepackage{ifpdf}
\input{Qcircuit}

\providecommand{\ket}[1]{|#1\rangle}
\providecommand{\bra}[1]{\langle#1|}

\providecommand{\idop}{\mathbbm 1}

\newcommand{\prlsection}[1]{{\em {#1}:--~}}

\newcolumntype{C}[1]{>{\centering\arraybackslash$}p{#1}<{$}}

\begin{document}
\title{Classical Computation by Quantum Bits}
\author{B. Antonio$^1$, J. Randall$^{2,3}$, W. K. Hensinger$^2$, G. W. Morley$^4$, S. Bose$^1$}
\affiliation{$^1$Department of Physics and Astronomy, University College London, Gower Street, WC1E 6BT London, United Kingdom}
\affiliation{$^2$Department of Physics and Astronomy, University of Sussex, Brighton, BN1 9QH, UK}
\affiliation{$^3$ QOLS, Blackett Laboratory, Imperial College London, London, SW7 2BW, UK}
\affiliation{$^4$Department of Physics, University of Warwick,
Gibbet Hill Road, Coventry CV4 7AL, United Kingdom}

\begin{abstract}
Atomic-scale logic and the minimization of heating (dissipation) are both very high on the agenda for future computation hardware. An approach to achieve these would be to replace networks of transistors directly by classical reversible logic gates built from the coherent dynamics of a few interacting atoms. As superpositions are unnecessary before and after each such gate (inputs and outputs are bits),  the dephasing time only needs to exceed a single gate operation time, while fault tolerance should be achieved with low overhead, by classical coding. Such gates could thus be a spin-off of quantum technology much before full-scale quantum computation. Thus motivated, we propose methods to realize the 3-bit Toffoli and Fredkin gates universal for classical reversible logic using a single time-independent 3-qubit Hamiltonian with realistic nearest neighbour two-body interactions. We also exemplify how these gates can be composed to make a larger circuit. We show that trapped ions may soon be scalable simulators for such architectures, and investigate the prospects with dopants in silicon. 
\end{abstract}

\maketitle
\prlsection{Introduction} Power dissipation has become a serious obstacle to packing more transistors per unit area so that the exponential rise of computational power with time (Moore's law) may be continued. Heating in a chip is projected to reach $200$Wcm$^{-2}$ by 2020~\cite{Lin2008} by the International Technology Roadmap for Semiconductors (ITRS). The search for less dissipative alternatives is on with suggestions such as molecular electronics~\cite{Lorente2013}, spin-wave computation~\cite{Kostylev2005}, magnetic and quantum dot cellular automata~\cite{Imre2006}, DNA logic~\cite{Khullar2007} and superconducting logic in cryogenic temperatures~\cite{Holmes2013}, which are by no means exhaustive. 

Aside from minimizing dissipation, another driving factor for contemporary computer technology is the atomic scale storage of information~\cite{Delgado2012,Loth2012,Lorente2013} and atomic scale logic~\cite{Fuechsle2012}. However, to our understanding, they do not yet aim to exploit the dynamics of highly isolated systems for computation in the same sense as the ``friction free" billiard ball computer of Fredkin and Toffoli~\cite{Fredkin1982}. Such dynamics is well approximated in the systems being developed for quantum technologies where quantum coherence is preserved by a high isolation. In fact, the energy dissipation time-scale $T_1$ can be exceptionally high, and even the dephasing time $T_2$ is fairly high. It is thereby worth studying whether the huge development towards quantum computation can, on the way to that grand goal, also provide a minimally dissipative, as well as miniaturized, hardware for classical logic. Any reliability sacrificed by going to atomic bits is not particularly new at this scale, and, is present even at the nano-scale, and ingenious ways of using low-reliability devices is topical~\cite{Anghel2007,Palem2013,Han2013}.


Motivated by the above, here we investigate whether an unmodulated minimal widget of 3 permanently interacting qubits, with each qubit encoding a microscopic bit, can act as a logic gate for classical computation. The aim is to (a) use coherent dynamics (to avoid dissipation and heating), (b) use permanent nearest neighbour two-body couplings of similar strength (to keep things realistic for a structure of proximal spin qubits, for example), (c) use a single ``time-independent" Hamiltonian to accomplish the entire gate and finally, (d) avoid any auxillary systems, hybrid systems or additional levels aside those of the relevant qubits. In practical implementations, additional regular pulsings may however be required for reducing decoherence through dynamical decoupling. We suggest placing these gates next to each other spatially to compose a classical circuit (we also exemplify this composability). The whole classical circuit will then simply be a 2D pattern of the widgets implementing the fundamental gates, where each such gate is implemented by a quantum evolution. However, the quantum state is allowed to decohere ``before" and ``after" each gate. This is acceptable for a classical circuit as the inputs and outputs of each gate are classical bits, and no superposition has to be maintained between the end of one basic gate and the start of another. Unlike the case for quantum computation, in the space of time between the gates, the correction of dephasing errors are not necessry. The use of classical codes therefore suffices to keep track of the errors within the computation.

As coherent dynamics is not only non-dissipative, but also reversible,  we will aim to build 3-bit Toffoli and Fredkin gates, which enable reversible classical computation~\cite{Bennett1973,Fredkin1982}. Reversible logic avoids heating due to the erasure of information~\cite{Landauer1961}. Quantum computation and quantum error correction has already motivated the implementation of Toffoli gates \cite{Cory1998,Monz2009,Fedorov2012}. Although these gates can be decomposed into 2-qubit gates and local unitaries~\cite{NielsenChuang}, such implementations require at least five 2 qubit gates~\cite{Yu2013}. To achieve simplifications beyond this remit (including single pulse or ``single-shot" implementations), nearly all schemes and actual implementations have variously used auxillary modes aside the relevant qubits, such as the cavity mode in hybrid qubit-resonator systems \cite{Chen2012,Chen2006,Xiao2007,Shao2013,Zheng2013,Fedorov2012} or the motional modes in ion traps \cite{Monz2009, Ivanov2011}, or auxillary levels outside the space of qubits \cite{Ralph2007,Zheng2013,Zahedinejad2015}. Where solely qubits have been used, either multiple pulses \cite{ivanov2015} or non-uniform and long-range couplings (departing from aim (b)) are generically present as in the NMR and other literature \cite{Du2001,Kumar2013}. Crucially, the key question of the ``possibility" and an analytic expression of the ``accuracy" of gates  in the simplest setting: 3 qubits and a time independent realistic nearest neighbour Hamiltonian, remains open. In fact, the conjuction (a)-(d) should be impossible as the generation of the unitary operations corresponding to the ideal Fredkin and Toffoli gates with a single (time-constant) 3 qubit Hamiltonian seems to necessitate unrealistic 3-body interactions \cite{chau1995,ionicioiu2009} $\sigma_i^z \bm{\sigma}_j \cdot \bm{\sigma}_k$ and $\sigma_i^z \sigma_j^z \sigma_k^x$ respectively. However, it has gone unnoticed that when we lower our aim from quantum to classical computation, i.e., when the relative phases between the computational basis states are irrelevant, and when approximate gates with low errors could be useful, then
classical Fredkin and Toffoli gates of useful accuracy become feasible with 3 qubit realistic nearest neighbour time-independent Hamiltonians.

\begin{figure}[b]
\begin{center}
\includegraphics[scale = 0.5]{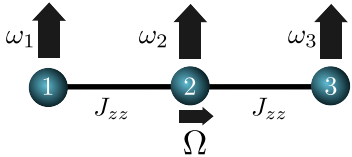}
\caption{Setup for creating a 3-qubit Toffoli gate.}
\label{fig:ZZ}
\end{center}
\end{figure}

 {\em Toffoli gate:-} A classical Toffoli gate flips the target bit when both the control bits are in the logical state $1$ (a quantum Toffoli is a unitary operator that additionally encodes a specific phase relationship between distinct quantum states). We start by describing how a structure of 3 permanently Ising coupled spins can be used to implement an {\em approximate} classical Toffoli gate and investigate how good the approximation can be. The gate is switched on by applying a transverse field to the target qubit (qubit 2 in this case, see fig.~\ref{fig:ZZ}). With this field switched on, the gate is performed through the time-independent Hamiltonian
\begin{equation}
H_{\textsc{tof} }= \frac{ J_{zz}}{2} ( \sigma_1^z\sigma_2^z + \sigma_2^z \sigma_3^z) +  \sum_j  \frac{\omega_j \sigma_j^z}{2}+  \frac{\Omega\sigma_2^x}{2} ,
\end{equation}
where $J_{zz}, \omega_j$ and $\Omega$ are in frequency units (unless otherwise specified we will use these units).  For the Toffoli gate we require  $\frac{1}{\sqrt{2}}\ket{1} (\ket{0} \pm e^{i\phi}\ket{1})\ket{1}$ to be eigenstates of $H_{\textsc{tof} }$. By applying $H_{\textsc{tof} }$ on states of this form, it can be confirmed that $\omega_2 = 2J_{zz}$ achieves the desired eigenstate with $\phi=0$, while $\omega_1$ and $\omega_3$ can remain arbitrary. The eigenstates and energies are then
\begin{align}
\ket{\pm}_{ij} &= \frac{1}{\mathcal{N}^{\pm}_{ij}}\text{ } \ket{i}  \left[(d_{ij} \pm \sqrt{1 + d_{ij}^2})\ket{0} + \ket{1} \right] \ket{j},\\
E_{ij}^{\pm} /\hbar &=  \frac{1}{2} \left[ (-1)^{i}\omega_1 + (-1)^{j} \omega_3  \pm \Omega \sqrt{d_{ij}^2 + 1} \right]
\end{align}
where $i,j \in \{0,1\}$, $d_{11} = 0$, $d_{01} = d_{10} = \omega_2/\Omega$, $d_{00} = 2\omega_2 /\Omega$, and $\mathcal{N}^{\pm}_{ij}$ are normalising factors. 
The $\ket{101} \leftrightarrow\ket{111}$ swapping will occur when $\exp\left[ -iH_{\textsc{tof} }t\right] \ket{101} = e^{i\theta} \ket{111}$, which occurs at a time $t = \tau_n = (2n+1)\pi \hbar /|E_{11}^{+} - E_{11}^{-}| = (2n+1)\pi / \Omega $ (where $n$ is an integer, and assuming without loss of generality that $\Omega$ is positive). In general, the evolution of the arbitrary computational basis states in this time $\tau_n$ is captured by the fidelities $f_{lmn \rightarrow xyz} := \bra{xyz} e^{-i H_{\textsc{tof}  }\tau_n } \ket{lmn}$.
As $[H_{\textsc{tof}  },\sigma_1^z]=[H_{\textsc{tof}  },\sigma_3^z]=0$ only the following fidelities are relevant
\begin{align}\label{eqn:Fid}
&f_{abc \rightarrow a \bar{b} c} =  \frac{-ie^{-i\phi_{ac}} (2n+1) \pi}{2}\mbox{sinc} \left( \frac{(2n+1)\pi}{2} \sqrt{d_{ac}^2 + 1} \right) \nonumber\\
&f_{abc \rightarrow abc} = -ie^{-i\phi_{ac}} \left[ \cos \left(\frac{(2n+1)\pi}{2} \sqrt{d_{ac}^2 + 1} \right) \right. \nonumber\\
& \left. - \frac{(2n+1)  i \pi}{2} d_{ac} \text{ }\mbox{sinc} \left( \frac{(2n+1)\pi}{2} \sqrt{d_{ac}^2 + 1} \right) \right]
\end{align}
where $a,b,c \in \{0,1\}$, $\bar{b} := b \oplus 1$, and 
$\phi_{ac} =\frac{(2n+1)\pi ((-1)^a \omega_1 + (-1)^c \omega_3)}{2\Omega}$.
Note that $|f_{101\leftrightarrow 111}| = 1$ by our choice of $\tau_n$.
To realise a Toffoli gate we further require that $|f_{i0j}\rightarrow f_{i1j}| = 0$ for $|ij\rangle \ne |11\rangle$ so that for these fidelities, the phase inside the sinc function in (\ref{eqn:Fid}) must be an integer multiple of $\pi$. This leads us to 
\begin{align}
\frac{1}{2} \sqrt{\frac{\omega_2^2}{\Omega^2} + 1} = \frac{m_1}{(2n+1)},~ \frac{1}{2} \sqrt{\frac{4\omega_2^2}{\Omega^2} + 1} = \frac{m_2}{(2n+1)},\label{eqn:Diop1}
\end{align}
where $m_1,m_2$ are non-zero integers, which in turn implies $
16 m_1^2 - 4m_2^2 = 3(2n+1)$, where
the left hand side is even, and the right side is odd. Thus no choice of $\omega_2,$ $\Omega$  gives a perfect Toffoli (a price to pay for the simplicity of $H_{\textsc{tof}}$).

However, we can find parameters which achieve  an approximate gate. Assuming $n=0$ for the shortest possible gate time and further assuming that the first subequation of Eq.(\ref{eqn:Diop1}) is exact with large $m_1$ one finds $m_2 \approx 2m_1$ from the second subequation of Eq.(\ref{eqn:Diop1}).
 With this choice of parameters, the phases inside the sinc functions in (\ref{eqn:Fid}) are all either  multiples of $\pi$ or approximately so up to order $1/m_1$. 
To evaluate how close this approximate Toffoli is to the exact Toffoli, we use the process trace distance, which for a 3-qubit system is defined as~\cite{Gilchrist2005} 
$\mathcal{D}_{pro} = \text{tr} \left| \chi(U) - \chi(T)\right|/16$
where $\chi(M)_{mn} = \mathrm{tr}(A_m^\dagger M)(\mathrm{tr}(A_n^\dagger M))^*$, and $\{ A_n \}_{n=1}^{64}$ is a complete orthogonal basis which satisfies $\mathrm{tr}(A_n^\dagger A_m) = \delta_{nm}$, 
$T$ is an ideal Toffoli gate, $U$ is the gate we can achieve with the above setup, and $|X|$ indicates the matrix norm. As we are interested in creating a ``classical" gate, we ignore any phases and define $U_{|f|}$ such that $ \bra{y}U_{|f|}\ket{x} = |f_{x \rightarrow y }|$ so this will only measure how close we are to a Toffoli gate apart from local operations. $\mathcal{D}_{pro}$ gives an upper bound on the average probability $\bar{p}_e$ that the gate fails~\cite{Gilchrist2005}, so
\begin{align}
\label{eqn:p}
\bar{p}_e \lesssim \mathcal{D}_{pro} = \frac{3 \pi}{16 m_1^2} + O\left( \frac{1}{m_1^4} \right)\approx \frac{3\pi}{4}\left(\Omega/\omega_2\right)^2 
\end{align}
In summary, we can achieve an approximate classical Toffoli gate with average failure error of Eq.(\ref{eqn:p}), if $J_{zz} = \omega_2/2$ and ${\omega_2}/{\Omega} =\sqrt{4 m_1^2 - 1}$, with $m_1$ large.

\label{sec:Heis}
\begin{figure}[b]
\includegraphics[scale = 0.5]{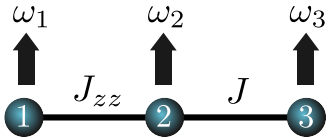}
\caption{Setup for creating a 3-qubit Fredkin gate, using Ising and Heisenberg coupling.}
\label{fig:FredHeis}
\end{figure}

{\em Fredkin Gate:-} We now consider creating a classical Fredkin gate (controlled-SWAP), using quantum Ising and  Heisenberg interactions. We consider the Hamiltonian 
\begin{align}\label{eqn:HamFred}
H_{\textsc{fred}} = \frac{J}{2}\bm{\sigma}_2 \cdot \bm{\sigma}_3 +\frac{ J_{zz}}{2} \sigma_1^z \sigma_2^z +\sum_j \frac{\omega_j \sigma_j^z}{2},
\end{align}
where qubit 1 is the control qubit, and qubits 2 and 3 are to be swapped (see fig.~\ref{fig:FredHeis}). The intuition is that the swapping induced by the Heisenberg interaction between qubits 2 and 3 will only occur when qubit 1 is in a state that makes the energy splitting of qubits 2 and 3 match.
For swap between qubits $2$ and $3$, we need to choose parameters such that states of the form $\frac{1}{\sqrt{2}} (\ket{110} \pm \ket{101} )$ are eigenstates of $H_{\textsc{fred}}$, which requires $J_{zz} = \omega_2 - \omega_3$. With these parameters, $\{ \ket{100},\ket{111},\ket{011},\ket{000} \}$ are all eigenstates, and the only eigenstates of the Hamiltonian which are not computational basis states are
\begin{eqnarray}
\ket{\psi}_{110}^{\pm} &= &(\ket{110} \pm \ket{101} )/\sqrt{2}, ~\ket{\psi}_{010}^{\pm} = \{J\ket{010}+(\omega_2 \nonumber\\
&&- \omega_3 \pm \sqrt{(\omega_2 - \omega_3)^2 + J^2}) \ket{001}\}/\mathcal{N}^{\pm}_{010} 
\end{eqnarray} 
where $\mathcal{N}^{\pm}_{010}$ is a normalising factor. The eigenenergies of these states are $E_{110}^{\pm} /\hbar= -\left( \omega_1 + J  \mp 2 J \right)/2$ and $E_{010}^{\pm}  /\hbar = \left( \omega_1 -  J \pm 2 \sqrt{J_{zz}^2 + J^2} \right)/2$ respectively.
The swap $\ket{110} \leftrightarrow \ket{101}$ is complete at a time $\tau_n = (2n+1) \pi \hbar/|E_{110}^+ - E_{110}^-| = (2n+1)\pi /2J$ (with $n \in \{0,1,2,...\}$, and assuming $J > 0$, without loss of generality). The fidelity for swapping $\ket{010} \leftrightarrow \ket{001}$ at time $\tau_n$ is $\propto \mbox{sinc} \left[(2n+1)\pi \sqrt{J_{zz}^2 + J^2} /J \right].$
For this fidelity to be zero, we need $J^2 \left(m^2/(2n+1)^2-1 \right) =(\omega_2 - \omega_3)^2$, where $m$ is an integer greater than 1, and $\frac{m}{(2n+1)} > 1$.

{\em Possible realizations:-} In implementations, the gate operation times, $\sim 1/\Omega$ for the Toffoli, and $\sim 1/J$ for the Fredkin, have to be smaller than the dephasing times. We discuss two possible realizations.

{\em Trapped Ions:-}
To engineer the Hamiltonian $H_{\textsc{tof}}$ we exploit the fact that an axial magnetic field gradient realises Ising couplings \cite{Wunderlich2001}. Consider three $^{171}\mbox{Yb}^+$ ions in a linear Paul trap with secular frequency  
$\nu = 2\pi \times 100  \mbox{kHz}$. The qubits are encoded in the $^2S_{1/2}$ $\ket{\downarrow} = \ket{F=0, m_{F}=0}$ and $\ket{\uparrow} = \ket{F=1, m_{F}=1}$ states which are separated by approximately $12.6 \text{GHz}$.  For $\partial_z B_j=250 \text{Tm}^{-1}$, one has  $J_{12} =J_{23} = J_{zz}= 2\pi \times 9.98 \mbox{kHz} $, $J_{13} =2\pi \times 7.07 \mbox{kHz}$. The extra $J_{13}$ coupling introduces extra phases to the gate but this has no effect if the gate is used for classical computation. The $\sigma_2^x$ field is achieved by applying a near-resonant microwave pulse leading to the trapped ion Hamiltonian 
\begin{align}\label{eqn:IonTrap1}
H^{(i)} &=\sum_{j=1}^3 \frac{\omega_j^0}{2} \sigma_j^z +  \frac{J_{zz}}{2}\sum_{j=1}^2 \sigma_j^z \sigma_{j+1}^z  + \Omega \cos( \omega_x t )\sigma_2^x,
\end{align}
where $\omega_j^0$ is the qubit energy splitting of ion $j$.
We can transform $H^{(i)}$ to a frame rotating with the operator $\frac{1}{2}(\sum_j \omega_j^0 \sigma_j^z -\delta\sigma_2^z)$, where $\delta$ is the detuning of the microwave field from the resonant frequency of ion 2. States in this rotating frame evolve according to 
\begin{align}
H^{(i)}_I = \frac{\delta}{2} \sigma_2^z + \frac{J_{zz}}{2}\sum_{j=1}^2 \sigma_j^z \sigma_{j+1}^z 
+  \Omega e^{i(\delta-\omega_2^0)\sigma_2^z t} \cos(\omega_x t )\sigma_2^x \nonumber
\end{align}
Choosing $\delta = 2J_{zz}$ and $\omega_x = \omega_2^0  - \delta$, and applying identity~\ref{eqn:Iden1} of the supplementary material, we have 
\begin{eqnarray}
H^{(i)}_{I} =  H_{\textsc{tof}} (\omega_2 = 2 J_{zz}, \omega_1=\omega_3=0) + O\left( \frac{\Omega}{2\omega_2^0  -4J_{zz}}\right).\nonumber 
\end{eqnarray}
Thus with $\Omega = 2 \pi  \times 1.8 \mbox{kHz} $ and $\omega_2^0 \sim 12.6 \text{GHz}$, $H^{(i)}_{I} \approx H_{\textsc{tof}}$, while the systematic gate error from Eq.(\ref{eqn:p}) is $\sim 0.02$. The gate time is $\sim 0.3\text{ms}$, giving an error due to decoherence of around $(1- e^{-t_{gate}/T_2})\sim 0.03$ with $T_2 \approx 10 \mbox{ms}$~\cite{Piltz2013} using dynamical decoupling). In addition there may be additional heating due to proximity of the ions to the electrode surface, which we estimate from the results in~\cite{Hughes2011}. For a cryogenically cooled surface trap with an ion-electrode distance of 160 $\mu$m we estimate an additional decoherence rate of 50 Hz, which results in an error of around 0.01. Overall we therefore expect an average gate with error $\sim 0.04$, which should be good enough for error-free classical circuitry \cite{Von1956,Dobrushin1977,Pippenger1991,Feige1994}.

\emph{Bismuth Donors in Silicon:-} 
We propose placing the donor atoms \cite{Morley2010,Mohammady2010,Mohammady2012,Balian2012,Morley2013,Wolfowicz2013,Morley2015} close to each other so that their electronic spins are permanently coupled by isotropic Heisenberg interactions. To engineer $H_{\textsc{tof}}$, the nuclear spins are prepared in different states~\cite{Morley2010,Kalra2014} $\{ I_1^z,I_2^z,I_3^z \}$, resulting in different hyperfine couplings (with the very high coupling strength $A=1.475$ GHz of Bi) at each site. Starting from a nuclear spin polarised sample \cite{Sekiguchi2010} we can flip the nuclear spin from $\frac{9}{2}$ to $-\frac{9}{2}$ in 9 steps of $\sim$10$\mu s$ each. The nuclear spins are stable for hours~\cite{Feher1959,Castner1962}, so this process could be done once before many operations of the gate. 
We use a magnetic field $\omega_L >> A$ to ensure the nuclear spin does not evolve. The Hamiltonian of the donors is then~\cite{Kane1998}
\begin{equation}
H^{(d)}_0=\omega_L\sum_{n=1}^3 S_n^z+\sum_{n=1}^3 A I_n^z S_n^z+\sum_{n=1}^2 J_{n,n+1} \bm{S}_n \cdot \bm{S}_{n+1},
\end{equation}
where $S_n^\alpha = \frac{1}{2}\sigma_n^\alpha$ with $\alpha = x,y,z$ are the Pauli matrices for electronic spin.
An AC field of strength $\Omega$ is applied on qubit 2 in the $x$-direction (which could be applied globally since the hyperfine splittings are different) so that the Hamiltonian
$H^{(d)} = H^{(d)}_0 + \Omega \cos ( \omega_x t )S_2^x$ acts on the donors.
 Setting $J_{12}=J_{23}=J,\omega_x = \omega_L + A I_2^z - J$ and transforming $H^{(d)}$ to a frame rotating with the operator $\omega_L\sum_{n=1}^3 S_n^z - JS_2^z + \sum_{n=1}^3 A I_n^z S_n^z$ results in (using identities~\ref{eqn:Iden1}-\ref{eqn:Iden2} of the supplementary material)
\begin{align}
&H^{(d)}_I = \frac{H_{\textsc{tof}}(J_{zz}=J, \omega_2=2J, \omega_1=\omega_3=0)}{2}\nonumber\\  
&+ O\left(\sum_{n=2}^3 \frac{J}{|A I_n^z - A I_{n-1}^z|} + \frac{\Omega}{2(\omega_L -2J + AI_2^z)} \right). \nonumber
\end{align}

%
Setting $I_1^z = \frac{9}{2}$, $I_2^z=-\frac{9}{2}$, and $I_3^z =\frac{9}{2}$, and with $\Omega = 1$MHz, $J = 30$MHz, both the error term in $H^{(d)}_I$ and the systematic gate error term in~Eq.(\ref{eqn:p}) are $\sim 10^{-3}$. The gate time is $2\mu s$ (thereby allowing the bandwidth of the AC pulse to significantly exceed the $\sim 2 \text{kHz}$ linewidths seen in experiments~\cite{Laucht2015}) so that
the errors due to decoherence are roughly $1-e^{-t_{gate}/T_2} \sim 10^{-6}$ (as $T_2 \sim $700ms in isotopically pure silicon~\cite{Wolfowicz2012}). 


%

The Fredkin gate can be implemented solely with $H^{(d)}_0$. Transforming $H^{(d)}_0$ to a rotating frame with the operator
$\omega_L \sum_{n=1}^3 S_n^z + A I_1^z S^z_1 + A I_2^z(S_2^z + S_3^z)$ and using the identity~\ref{eqn:Iden2} of the supplementary material gives
\begin{align}
&&H^{(d)}_{0,I} = \frac{1}{2}H_{\textsc{fred}}(J_{zz}=J_{12},J=J_{23},\omega_2 \nonumber\\&-& \omega_3 = 2A (I_2^z - I_3^z))+O\left( J_{12}/|A I_1^z - A I_2^z|\right).
\label{error-fred}
\end{align}
Setting $ I_1^z = -\frac{9}{2}, I_2 = \frac{9}{2}, I_3 = \frac{7}{2}$, Fredkin gate conditions $J_{12} = 2(A I_3^z - A I_2^z)=J_{23}\sqrt{\frac{m^2}{(2n+1)^2}-1} $ can be met by $J_{23} = J_{12}(1+10^{-6} )$ ($n = 675,m = 2340$) with a resulting gate time $\sim 0.5 \mu\text{s} = 10^{-3}T_2$ so that decoherence is negligible. The error term in Eq.(\ref{error-fred}) is $\sim 0.22$, but could be minimized further to $\sim 0.07$ by techniques mentioned in the supplementary material.

\begin{figure}[h]
\includegraphics[scale = 0.4]{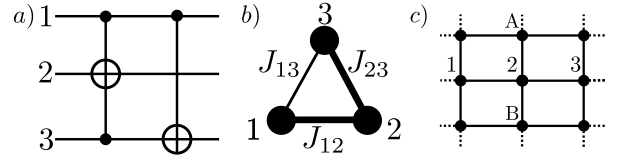}
\caption{a) A half-adder circuit. b) Setup for creating a half adder, using two pulses and with $J_{12} = J_{23} \neq J_{13}$. c) Using selective addressing on arrays of qubits, general computations can be achieved.}
\label{fig:HalfAdder}
\end{figure}

\emph{ Composability:-} To exemplify circuit building, we show how a half-adder (fig.~\ref{fig:HalfAdder} a)) \cite{Roy2015} can be implemented with the arrangement of Ising coupled spins shown in fig.~\ref{fig:HalfAdder} b). The Toffoli gate with qubit 2 as target is implemented using $H_{\textsc{tof}}$. Then we want to apply the controlled-\textsc{not} irrespective of the state of qubit 2. Two successive pulses on qubit 3 of frequencies $\omega_+$ and $\omega_-$, where $\omega_{\pm} =  \omega_3 -J_{13} \pm J_{23}$ implement two successive conditional flips of qubit 3 according to when qubits 1 and 2 are in a $\ket{10}_{12}$ and $\ket{11}_{12}$ respectively. Each of these pulses implement different time-independent Hamiltonians in appropriate rotating frames. Thus, 3 successive time independent Hamiltonians implement a half adder. In general, a Toffoli gate can be applied on a set of qubits (say, 1, 2 and 3 with 2 as target) of the Ising coupled array depicted in fig.~\ref{fig:HalfAdder} c), by pulses of 4 frequencies to flip qubit 2 irrespective of the state of those neighbours (say, A and B) that do not take part in the gate. While our pulsing is similar to tools in liquid state NMR \cite{Du2001,Roy2015}, it was not apparent to date that the gates possible in the simplest of settings are approximate, inequivalent to the unitary operations corresponding to Fredkin and Toffoli gates (to achieve those further ``non-local" gates are necessary), and do not require long-range couplings.

%
%

{\em Reliability:-} Our physical realizations have errors (as in any nano-scale logic, including scaled CMOS). However, reliable classical computation with faulty components is possible with a constant overhead ~\cite{Von1956,Pippenger1991} as our error rates are below the $1/6$ required threshold~\cite{Hajek1991}. One can use measurements between gates and a classical resetting of bits (e.g., parity protected gates~\cite{Parhami2006}). Besides, applications such as image processing tolerate more noise~\cite{Han2013}. If we automate error correction
with faulty gates such as in ``quantum" error correction, thresholds $\sim 10^{-2}-10^{-3}$ are however obtained (see Supplementary Material~\ref{app:FT}), which are still met by the donor based implementation of $H_{\textsc{tof}}$.

{\em Conclusions:-} We have demonstrated how the classical Toffoli and Fredkin gates can be achieved by realistic 3 qubit time-independent Hamiltonians. This focus on simplicity (e.g., time-independence, no auxillary systems/levels) as opposed to fidelity \cite{Zahedinejad2015}, stems from aiming to build low dissipation atomic-scale classical logic. Targeting classical gates helps us circumvent the apparent impossibility of the ideal Toffoli and Fredkin unitaries under our desiderata (a)-(d).  Although AC pulses were used in the proposed implementations with trapped ions and Bi donors, and for building circuits, these were merely a means to implement the time-independent $H_{\textsc{fred}}$ and $H_{\textsc{tof}}$ in appropriate rotating frames. This only results in extra relative phases between the computational basis states that do not matter for classical computing.  While our gates minimize the dissipation in computation, the study of dissipation while pulsing and measuring and the possibility of room temperature realizations \cite{Yao2012} is kept for the future.
 
\acknowledgements

SB is supported by the European Research Council under the European Union’s Seventh Framework Programme (FP/2007-2013) / ERC Grant Agreement n. 308253. GWM is supported by the Royal Society. This work is supported by the U.K. Engineering and Physical Sciences Research Council  [EP/G007276/1, the UK Quantum Technology hub for Networked Quantum Information Technologies (EP/M013243/1), the UK Quantum Technology hub for Sensors and Metrology (EP/M013243/1)], the European Commission’s Seventh Framework Programme (FP7/2007-2013) under Grant Agreement No. 270843 (iQIT), the Army Research Laboratory under Cooperative Agreement No. W911NF-12-2-0072, the US Army Research Office Contract No. W911NF-14-2-0106 and the University of Sussex. The views and conclusions contained in this document are those of the authors and should not be interpreted as representing the official policies, either expressed or implied, of the Army Research Laboratory or the U.S. Government. The U.S. Government is authorized to reproduce and distribute reprints for Government purposes notwithstanding any copyright notation herein.

%


\appendix

\section{Interaction picture and the rotating wave approximation}
\label{app:IntPic}

Consider a Hamiltonian that contains some fast oscillating terms of the form $V\cos \omega t$, where $V$ is time independent and Hermitian. If $\omega > \| V \|$, then the evolution operator can be approximated by (see e.g.~\cite{DalessandroBook})
\begin{align}
U(t,0) = \idop + O \left( \frac{\Vert V \Vert}{ \omega} \right) 
\end{align}
To represent this, we will use the notation $H(t) = H_0(t) + O \left( \frac{\Vert V \Vert}{ \omega} \right)$. We now derive two identities that are useful in the main text. For this first identity, consider a Hamiltonian term of the form $C  e^{-i\omega \sigma^z t/2} \cos (\omega_x t) \sigma^x e^{i\omega \sigma^z t/2}$. Setting $\omega_x = \omega$ gives
\begin{align}\label{eqn:Iden1}
& C e^{-i\omega \sigma^z t/2} \cos (\omega t) \sigma^x e^{i\omega \sigma^z t/2}\nonumber \\
&=C e^{-i\omega \sigma^z t} \cos (\omega t) \sigma^x=\frac{C}{2} e^{-i\omega \sigma^z t} \sigma^x (e^{i\omega t} +  e^{-i\omega t}) \nonumber\\
&=\frac{C}{2}(e^{-i\omega t } \sigma^+ + e^{i\omega t } \sigma^- )(e^{i\omega t } +  e^{-i\omega t})  \nonumber\\
&= \frac{C}{2}\left(  \sigma^+ + e^{2 i\omega t } \sigma^- +  e^{-2i\omega t} \sigma^+ +  \sigma^- \right)\nonumber\\
& = \frac{C}{2} \sigma^x + O \left( \frac{C}{ 2\omega} \right)
\end{align}
provided that $C <   2\omega$. For the second identity, consider a Hamiltonian of the form $e^{-it [\omega_1 \sigma^z_1 +  \omega_2 \sigma^z_2]/2}J( \sigma^x_1 \sigma^x_2 + \sigma^y_1 \sigma^y_2 + \sigma^z_1 \sigma^z_2) e^{it [\omega_1 \sigma^z_1 +  \omega_2 \sigma^z_2]/2}$ 
\begin{align}\label{eqn:Iden2}
&Je^{-it [\omega_1 \sigma^z_1 +  \omega_2 \sigma^z_2]/2} (\sigma^x_1 \sigma^x_2 + \sigma^y_1 \sigma^y_2 + \sigma^z_1 \sigma^z_2)e^{it [\omega_1 \sigma^z_1 +  \omega_2 \sigma^z_2]/2} \nonumber \\
&=2J e^{-it [\omega_1 \sigma^z_1 +  \omega_2 \sigma^z_2]} (\sigma^+_1 \sigma^-_2 + \sigma^-_1 \sigma^+_2) + J\sigma^z_1 \sigma^z_2 \nonumber\\
&=2J\sigma^+_1 \sigma^-_2 e^{it[\omega_1 - \omega_2]}  + 2J\sigma^-_1 \sigma^+_2 e^{-it[\omega_1 - \omega_2]} + J\sigma^z_1 \sigma^z_2 \nonumber \\
&=  J\sigma^z_1 \sigma^z_2 + O \left( \frac{4J}{\omega_1 - \omega_2} \right)
\end{align}
provided $ 4J < (\omega_1 - \omega_2)$.

\section{Thresholds for fault-tolerant classical computation}\label{app:FT}

We will estimate the error threshold required to implement a fault-tolerant classical Toffoli gate, using the simplest classical code, the 3-bit repetition code. We will use a similar analysis as in~\cite{NielsenChuang,Knill1998}, which is not a very rigourous analysis but will give a rough idea of the kind of classical threshold we will need to achieve with the Toffoli gate. A simple classically fault-tolerant Toffoli circuit can be constructed as follows:
\[
\Qcircuit @C=1em @R = 0.6em @!R  {
                  & \ctrl{5}    &    \qw      &     \qw        & \qw     & \multigate{2}{\mathcal{S}}  & \qw & \qw & \multigate{2}{\mathcal{R}} & \qw  \\
\lstick{C_1}&  \qw       &   \ctrl{5}   & \qw           & \qw     & \ghost{\mathcal{S}}             & \qw & \qw&  \ghost{\mathcal{R}}  & \qw  \\
                  &  \qw       &   \qw       &   \ctrl{5}     & \qw     & \ghost{\mathcal{S}}             & \qw& \qw &  \ghost{\mathcal{R}}   & \qw \\
                  &               &                 &                 &             & \mbox{Syndrome}              &         &       &  \mbox{Recovery}       &          \\
&&&&&&&&&\\
                  & \ctrl{5}  &  \qw        &  \qw          & \qw       & \multigate{2}{\mathcal{S}}  & \qw & \qw& \multigate{2}{\mathcal{R}} & \qw  \\ 
\lstick{C_2}&  \qw      & \ctrl{5}    &  \qw           & \qw      & \ghost{\mathcal{S}}             & \qw & \qw&  \ghost{\mathcal{R}}         & \qw \\
				  &  \qw      &  \qw        &  \ctrl{5}      & \qw      & \ghost{\mathcal{S}}             & \qw & \qw&  \ghost{\mathcal{R}}         & \qw \\
				  &               &                 &                 &             & \mbox{Syndrome}              &         &       &  \mbox{Recovery}       &          \\
&&&&&&&&&\\
               & \targ{X} & \qw         &  \qw          &\qw      & \multigate{2}{\mathcal{S}}       & \qw  & \qw& \multigate{2}{\mathcal{R}} & \qw   \\
\lstick{T} & \qw       & \targ{X}   &  \qw          & \qw     & \ghost{\mathcal{S}}                 & \qw  & \qw&  \ghost{\mathcal{R}}         & \qw  \\
               & \qw       & \qw         & \targ{X}     & \qw          & \ghost{\mathcal{S}}             & \qw & \qw &  \ghost{\mathcal{R}}         & \qw \\
				  &               &                 &                 &             & \mbox{Syndrome}                 &       &        &  \mbox{Recovery}       &          \\
&&&&&&&&&\\
%
}
\]

Where the syndrome + recovery step is

\[
\Qcircuit @C=1em @R = 0.6em @!R  {
& \multigate{2}{\mathcal{S}}  & \qw  & \qw & \multigate{2}{\mathcal{R}} & \qw  &   &  &\qw   &\ctrl{4}  & \ctrl{5}   & \qw       & \targ{X}  & \qw  & \qw & \qw\\
& \ghost{\mathcal{S}}            & \qw  & \qw &  \ghost{\mathcal{R}}          & \qw  & = & &\qw  & \ctrl{3}  & \qw       & \ctrl{5}   & \qw  & \targ{X} & \qw & \qw\\
& \ghost{\mathcal{S}}            & \qw  & \qw &  \ghost{\mathcal{R}}          & \qw &     & &\qw   & \qw      &  \ctrl{3}  & \ctrl{4}   & \qw  & \qw  & \targ{X}& \qw\\
& \mbox{Syndrome}              &         &        &  \mbox{Recovery}              &         &     &   &       &            & 			    & 				  & 		  & 		  & 		& 		\\
&&&&&&&&\lstick{\ket{0}}       & \targ{X}  & \qw       & \qw      & \ctrl{-4}  & \ctrl{-3}  & \qw         & \qw\\
&&&&&&&& \lstick{\ket{0}}       & \qw       & \targ{X}  & \qw      & \ctrl{-5} & \qw         & \ctrl{-3}  & \qw\\
&&&&&&&&\lstick{\ket{0}}        & \qw       & \qw        & \targ{X} & \qw      &  \ctrl{-5}  & \ctrl{-4}  & \qw\\
&&&&&&&&&&&&&&&\\
}
\]
Since the code is robust against single classical (bit-flip) errors, a rough expression for the fault tolerant gate threshold can be found by considering the ways in which 2 or more errors can occur. Two errors can be output by a single fault-tolerant Toffoli gate in the following ways:
\begin{itemize}
\item[1]There are incoming errors in two or more of the inputs, which can propagate through the gate to create two errors on one encoded qubit. Incoming errors would be from a single error exiting from a previous syndrome / recovery step, which could happen at any of six points in the syndrome / recovery process. There are $^3C_2 = 3$ ways that two errors could be incoming, so if the error of a bit flip in the gate is $p$ the total probability is $3 \times (6 p)^2 = 108 p^2$. 
\item[2] There is an incoming error in one of the inputs, and an error in the first round of Toffoli gates. Incoming errors have probability $6p$, and the probability of one error in any of the three gates is $3p$. There are 3 ways this can happen so the overall probability is $3 \times 3p \times 6p=56p^2$.
\item[3] Two errors occur during the Toffoli gates. There are $^3C_2 = 3$ ways this can happen, so the total probability is $3p^2$.
\item[4] One failure occurs during the Toffoli gates (three ways this can happen), and one during one of the syndrome gates (3 ways this can happen). Total probability for all three encoded qubits is $3 \times 3p \times 3p = 27 p^2$. 
\item[5] Two failures occur during one of the syndrome gates ($^3C_2$ ways this can happen per encoded qubit). Total probability is $3 \times 3p^2 = 9p^2$.
\item[6] One failure occurs during the syndrome gate and one during the recovery. Total probability $3 \times 3p \times 3p = 27p^2$.
\item[7] Two or more failures during recovery; total probability $9p^2$.
\end{itemize}
In total, this gives a rough probability of $(108 + 56 + 3 + 27 + 9 + 27 + 9)p^2 = 239p^2$ of two errors occurring undetected in the fault-tolerant Toffoli gate. Thus following concatenation of this encoding (see e.g.~\cite{NielsenChuang}) we would expect errors in the gate below the threshold $p \lesssim \frac{1}{239} \simeq 5 \times 10^{-3}$ to give a fault tolerant Toffoli gate.

\section{Parameter set for Fredkin Gate with Bismuth donors}
To satisfy the conditions of the Fredkin gate, $J_{12} = 2(A I_3^z - A I_2^z)$ and $J_{23}^2 (\frac{m^2}{(2n+1)^2}-1) =4(A I_3^z - A I_2^z)^2$.  For the minimal gate time ($n=1$), this would give $J_{23} = \frac{1}{\sqrt{3}} J_{12}$, or alternatively $J_{23}$ could be tuned to different fractions of $J_{12}$ by altering $m$ and $n$. For example, we can achieve $J_{23} = J_{12}(1+10^{-6} )$ by setting $n = 675$ and $m = 2340$. The resulting gate time would be $675\times 0.5\text{ns} \sim 0.5 \mu\text{s} = 10^{-3}T_2$, so such tuning would still not lead to large decoherence errors. 

\begin{figure}[h]
\includegraphics[scale = 0.5]{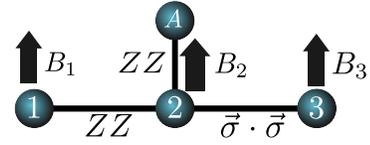}
\caption{Adding a fourth qubit to the original setup, in order to relax the constraints on $J_{zz}$ and provide a control to turn the gate on and off.}
\label{fig:FredHeisT}
\end{figure}

With $J_{12}=2(A I_3^z - A I_2^z)$, the error term in $H^{(d)}_{0,I}$ is $O(  2(I_2^z - I_3^z) / (I_1^z -  I_2^z))$, which is minimal when $ I_1^z = -\frac{9}{2}, I_2 = \frac{9}{2}, I_3 = \frac{7}{2}$. To decrease this error, a fourth qubit could be included, as shown in Fig.~\ref{fig:FredHeisT}. Adding this qubit $E$ with Ising coupling $J_{E2}$ effectively adds another local magnetic field to qubit 2 and so changes the resonance condition to $J_{12} +J_{E2}= \omega_3 - \omega_2$ (assuming that this qubit is set in the $\ket{0}$ state). Note however that this coupling also has an error associated to it, and so the optimal situation is with $J_{E2} = J_{12} =(A I_3^z - A I_2^z) = 1.475$GHz. This results in an error of around 11\%. Adding yet another control qubit $E'$ with coupling $J_{E'2}$ and choosing couplings such that $J_{12} = J_{E2} = J_{E'2} = \frac{2}{3}(A I_3^z - A I_2^z)$ results in errors of around 7\% (adding any more becomes unrealistic as the control qubits may begin to interact significantly). The gate time in this situation would be $\sim1.5\text{ns}$ which is still significantly smaller than $T_2$.

Additionally, setting $J_{12} = (\omega_2 - \omega_3)$ is not straightforward, since the donor position is not continuously tunable. There are also oscillations in the exchange interaction that depend on the separation between donors and the orientation of the donors relative to the crystal~\cite{Koiller2001}, although these oscillations can be minimised if donors are aligned along the [100] axis, or if strain is applied~\cite{Koiller2002}. Atomically-precise positioning of the donors is possible~\cite{OBrien2001,Schofield2003,Fuechsle2012}, so a possible approach is to position the donors at separations of around 15-20nm such that $J_{12} \approx (\omega_2 - \omega_3)$ and then use a magnetic field gradient to tune $\omega_2 - \omega_3$ closer to $J_{12}$ (this magnetic field would also decrease the error in $H^{(d)}_{0,I}$). Since it would be possible to bring a magnetic tip close to the sample, large magnetic field gradients of up to $10^7 \text{ Tm}^{-1}$ would be possible~\cite{Mamin2007} so that $\omega_2 -\omega_3$ could be tuned by up to $\pm 0.5\text{ GHz}$. Alternatively, we could adopt the method in~\cite{Kane1998}, and use electric gates to tune the inter-donor couplings and increase hyperfine interactions, however this might introduce extra noise due to charge fluctuations.

We could go further and use an extra qubit as an on/off switch for the gate, which could be useful if we wish to concatenate several of these gates together. Consider the setup in Fig.~\ref{fig:FredHeisT} that has one additional external qubit, labelled qubit E. For qubit E to act as a control, we just need to make sure that the resonances of qubits 2 and 3 only match when qubits 1 and E are in the $\ket{1}$ state, and are very different otherwise so that the Heisenberg coupling becomes effectively an Ising coupling. By finding the resonance energy of qubit 2 under different settings of qubits 1 and E, and choosing $J_{12} = J_{2E}$, we find that the following conditions must be satisfied 
\begin{align}
&A +2 J_{12} \gg J_{23},\text{       }A \gg J_{23},\text{	       }A - 2J_{12} \simeq 0
\end{align}
Altogether these mean that the conditions under which this on/off switch would work is $A \simeq J_{12}$, and $A, J_{12} \gg J_{23}$.

\end{document}